\documentclass [12pt]{article}
\usepackage{graphicx,epsfig}
\title{Charge and magnetic order in the spin-one-half Falicov-Kimball 
model with Hund coupling in two dimensions}
\author{Hana \v Cen\v carikov\'a, Pavol Farka\v sovsk\'y, Nat\'alia 
Toma\v sovi\v cov\'a\\ and Martin \v Zonda\\
Institute of Experimental Physics, Slovak Academy of Sciences\\
Watsonov\'a 47, 040 01 Ko\v sice, Slovakia}
\date{}
\begin{document}
\baselineskip=20pt
\maketitle

\begin{abstract}
The spin-one-half Falicov-Kimball model with spin-dependent on-site 
interaction between localized ($f$) and itinerant ($d$) electrons  
is studied by small-cluster exact-diagonalization calculations and a 
well-controlled approximative method in two dimensions. 
The results obtained are used to categorize the ground-state configurations 
according to common features (charge and spin ordering) for all $f$ and $d$
electron concentrations ($n_f$  and $n_d$) on finite square lattices. 
It is shown that only a few 
configuration types form the basic structure of the charge phase diagram in the 
$n_f-n_d$ plane. In particular, the largest regions of stability correspond 
to the phase segregated configurations, the axial striped configurations and 
configurations that can be considered as mixtures of chessboard configurations 
and the full (empty) lattice. Since the magnetic phase diagram is much richer 
than the charge phase diagram, the magnetic superstructures are examined only 
at selected values of $f$ and $d$ electron concentrations.
\end{abstract}

\newpage
%%%%%%%%%%%%%%%%%%%%%%%%%%%
\section{Introduction}
%%%%%%%%%%%%%%%%%%%%%%%%%%%%
The interplay between charge and spin degrees of freedom in strongly correlated 
systems has triggered enormous interest in recent years due to the rich variety 
of charge and spin orderings found in some rare-earth and transition-metal 
compounds. Charge and spin superstructures have been observed, for example, 
in doped niclate~\cite{Ni}, cuprate~\cite{Cu} and cobaltate~\cite{Co} materials,
some of which constitute materials that exhibit high-temperature 
superconductivity. One of the simplest models suitable to describe 
charge-ordered phases in interacting electron systems is the Falicov-Kimball 
model (FKM)~\cite{FKM}. Indeed, it was shown that the simplest version of this 
model (the spinless FKM) already 
exhibits an extremely rich spectrum of charge ordered solutions, including
various types of periodic, phase-separated and striped phases~\cite{Lemanski, Farky}. 
However, the
spinless version of the FKM, although nontrivial, is not able to account 
for all aspects of real experiments. For example, many experiments show that 
a charge superstructure is accompanied by a magnetic superstructure~\cite{Ni,Cu,
expspin}. 
In  order to describe both types of ordering in the unified picture, a simple 
model based on a generalization of the spin-one-half FKM with an anisotropic, 
spin-dependent interaction that couples the localized and itinerant subsystems 
was proposed~\cite{Lemanski2}. Thus the model Hamiltonian can be written as 
\begin{equation}
H=\sum_{ij\sigma}t_{ij}d^+_{i\sigma}d_{j\sigma} 
+ U\sum_{i\sigma\sigma'}f^+_{i\sigma}f_{i\sigma}d^+_{i\sigma'}d_{i\sigma'}
+ J\sum_{i\sigma}(f^+_{i-\sigma}f_{i-\sigma} - 
f^+_{i\sigma}f_{i\sigma})d^+_{i\sigma}d_{i\sigma}\ ,
\end{equation}

where $f^+_{i\sigma}, f_{i\sigma}$ are the creation and annihilation operators 
for an electron of spin $\sigma=\uparrow, \downarrow$ in the localized state at 
lattice site $i$ and $d^+_{i\sigma}, d_{i\sigma}$ are the creation and annihilation
operators of the itinerant electrons in the $d$-band Wannier state at site~$i$.

The first term of (1) is the kinetic energy corresponding to quantum-mechanical 
hopping of the itinerant $d$ electrons between sites $i$ and $j$. These intersite
hopping transitions are described by the matrix elements $t_{ij}$, which are 
$-t$ if $i$ and $j$ are the nearest neighbours and zero otherwise (in the following 
all parameters are measured in units of $t$). The second term represents the 
on-site Coulomb interaction between the $d$-band electrons with density 
$n_d=N_d/L=\frac{1}{L}\sum_{i\sigma}d^+_{i\sigma}d_{i\sigma}$ and the localized
$f$ electrons with density  
$n_f=N_f/L=\frac{1}{L}\sum_{i\sigma}f^+_{i\sigma}f_{i\sigma}$, where $L$ is the 
number of lattice sites. The third term is the above mentioned anisotropic, 
spin-dependent local interaction of the Ising type between the localized 
and itinerant electrons that reflects the Hund's rule force. Moreover, it is 
assumed that the on-site Coulomb interaction between $f$ electrons is infinite 
and  so the double occupancy of $f$ orbitals is forbidden.

Since the $f$-electron occupation number $f^+_{i\sigma}f_{i\sigma}$ of each site $i$ still 
commutes  with the Hamiltonian (1), the $f$-electron occupation number is a 
good quantum  number, taking only two values: $w_{i\sigma}=1$ or 0, according
to whether or not the site $i$ is occupied by the localized $f$ electron.
Now the Hamiltonian (1) can be written as 
\begin{equation}
H=\sum_{ij\sigma}h_{ij}d^+_{i\sigma}d_{j\sigma},
\end{equation}
where $h_{ij}=t_{ij}+(Uw_{i}+Jw_{i-\sigma}-Jw_{i\sigma})\delta_{ij}$. Thus 
for a given $f$-electron configuration $w=\{w_1,w_2,\dots,w_L\}$, the Hamiltonian 
(2) is the second-quantized version of the single-particle Hamiltonian $h(w)$, 
so the investigation of the model (2) is reduced to the investigation of the 
spectrum of $h$ for different configurations of $f$ electrons. This can be 
performed exactly, over the full set of $f$-electron distributions 
(including their spins), or approximatively, over the reduced set of 
$f$-electron configurations. Here we use a combination of both 
numerical methods. The same procedure has been used already in our previous 
paper~\cite{Farky1} to study the ground-states of the extended FKM in one dimension. 
In the mentioned paper we have also presented some preliminary results concerning the 
ground states of the model in two dimensions. These results revealed that the 
extended FKM with spin-dependent interaction between $f$ and $d$  electrons can 
describe various types of charge and magnetic superstructures. In the present
paper we try to construct the comprehensive phase diagram (in the $N_f-N_d$ 
plane) of the extended FKM in two dimensions. We supply our 
studies by a detailed finite-size scaling analyses in order to minimize the 
influence of finite-size effects on the ground-state properties of the model. 
Here we consider only the case of strong Coulomb interactions ($U=4$), since the 
one-dimensional studies showed that the influence of the Hubbard interaction 
(between $d$ electrons) on ground-states of the model Hamiltonian (2) can be 
neglected in this limit. At the end of this paper we specify precisely conditions 
under which this term can be neglected in two dimensions.

%%%%%%%%%%%%%%%%%%%%%%%%%%%%%%%%%%
\section{Results and discussion}
%%%%%%%%%%%%%%%%%%%%%%%%%%%%%%%%%%
To construct the comprehensive picture of charge and magnetic ordering in the 
extended FKM in two dimensions, the complete phase diagram of the model in 
the $N_f-N_d$ plane has been calculated  point by point for all even number 
of $N_f$ and $N_d$. Of course, such a procedure demands a considerable amount 
of CPU time, that imposes severe restrictions on the size of clusters that can 
be studied numerically ($L=8\times 8$). First we have concerned our attention 
on the problem of charge ordering. In Fig.~1 we present results of our numerical 
calculations obtained for $U=4$ and $J=0.5$ in the form of the skeleton phase 
diagram. 
One of the most interesting observations is that the phase diagram consists 
of only a few configuration types, 
although the total number of possible configurations increases very rapidly 
with the cluster size $L$ as $3^L$. In particular, we have detected 5 
different charge
configuration types, and namely: (a) the segregated configurations,
where $f$ electrons clump together, (b) the n-molecular phases, which have 
been observed only for small $N_d$ and small $N_f$, (c) the axial stripes, 
(d) the regular phases (stable only in isolated points) and mixtures of 
regular (usually chessboard) phases 
and empty/full lattice accompanied by (e) the miscellaneous 
configurations. 
The typical examples corresponding to these configuration types
are depicted in the lower part of Fig.~1. 
As one can see the largest stability regions correspond only three configuration 
types, and namely, the segregated configurations and the axial stripes, 
which fill the left and right side of the phase diagram and the regular 
phases/mixtures of regular phases and empty (full) lattice localized at the 
central part of the skeleton phase diagram. 
Moreover, it was found that  
the homogeneous $f$-electron distributions ($f$ electrons are distributed as 
far away from each other as possible) are the ground states only in the region 
$d$ and only in isolated points, while in the rest part of the phase diagram 
the inhomogeneous distributions (the segregated phases, the axial stripes, 
the n-molecular phases) are preferred. 
This result is very valuable since in the past years a strong interest in 
the 
experimental and theoretical studies of strongly correlated systems was 
focused on the physics that leads to an inhomogeneous ordering, especially to an 
inhomogeneous charge stripe order due to the observation of such 
ordering in doped niclate and cuprate materials some of which exhibit 
high-temperature superconductivity. From the theoretical point of view, the 
presented skeleton phase diagram clearly demonstrates that this relatively 
simple model (generalized two-dimensional FKM) can describe such 
inhomogeneous stripe ordering. 
Moreover, it was shown that the stability area of stripes is relatively large 
and, in addition, includes both physically interesting conditions 
$N_f+N_d=L$ and $N_f+N_d=2L$.

Let us now turn our attention on the second problem, and namely, the problem 
of spin ordering. We have found, that although the 
charge phase diagram of the generalized two-dimensional FKM is rather simple, 
the spectrum of magnetic solutions is very rich. 
As demonstrated in Fig.~2, the one charge distribution could have many 
different spin configurations and therefore the exact classification is very 
difficult.  Moreover, in the two-dimensional case the finite
size effects on spin orderings are still large for clusters with $L\leq 64$, 
and thus we were not able to construct definitive
picture of magnetic phase diagram for all $N_f$ and $N_d$. For this reason
we have focused our attention on the physically the most interesting cases 
mentioned above, and namely, $N_f+N_d=L$ and $N_f+N_d=2L$, where exhaustive studies have been performed. 
To minimize the 
finite size effects we have studied the set of finite clusters of $L=6\times 6$, 
$8\times 8$,  $10\times 10$, $12\times 12$, $16\times 16$, $18\times 18$ and  
$20\times 20$ sites. 

We have started our  study with the 
case $N_f+N_d=2L$, which is slightly simple for a description. 
As shown in Fig.~3 the ground states for $N_f+N_d=2L$ are  
antiferromagnetic (AF) with alternating pattern, where the electrons 
(for $n_f<1/2$) or holes (for $n_f>1/2$) form the axial distributions.
Our calculations showed that these inhomogeneous stripe distributions are 
stable for large Coulomb interactions, while decreasing $U$ leads to their
destruction and prefers the homogeneous electron arrangement (see Fig.~4). 

From the skeleton phase diagram we know that the condition line 
$N_f+N_d=L$ lies in the axial stripe area, but in comparison with the previous 
case the situation is fully different. The first fundamental difference is that 
for sufficiently small $f$-electron concentrations $n_f$ the ground state 
could be ferromagnetically (F) ordered (see Fig.~5). The second one is, that 
although with increasing $f$-electron concentration the ground states are 
the AF, these AF arrangements are formed by F ordered clusters (domains). 
In  addition, for $n_f=1/4$ and $n_f=1/3$ ($L\geq 144$) a new type of stripes 
(known as the ladders) is occurred. And finally, a detailed analyses showed that 
there exists a critical $f$-electron concentration $n_f^c\sim 1/4$ bellow 
which the ground-states are phase separated.

To verify the ability of our method to describe ground-state properties of
macroscopic systems we have performed similar calculations for $U=8$ and several
values of $n_f$ and $n_d$ ($n_f=3/4$, $n_d=1/2$; $n_f=2/3$, $n_d=2/3$; 
 $n_f=1/2$, $n_d=1$; $n_f=1/3$, $n_d=4/3$; $n_f=1/4$, $n_d=3/2$) for which 
there exist the numerical results
obtained in the thermodynamic limit ($L\to \infty$) using the method of
restricted phase diagrams~\cite{Lemanski2}. The numerical results for 
ground-state
configurations obtained by these two different approaches are compared in
Fig.~6. It is seen that both methods yield the same results for the charge
distributions of $f$-electrons, but differ in a prediction of spin
distributions. To check the stability of our solutions 
we have performed
numerical calculations also for $L=12\times 12, 16\times 16$ and $18\times
18$ clusters. The results obtained are shown in Fig.~7 and they clearly
demonstrate that ground states found for $L=8\times 8$ cluster hold also on
much higher clusters. This fact allows us to extrapolate our numerical
results to the thermodynamic limit, where they can be directly compared 
with the Lemanski's results. This comparison shows (see Fig.~8) that our
method yields in all cases the lower energy in the thermodynamic limit ($L\to
\infty$) than the Lemanski's approach. Moreover, a comparison of our
and Lemanski's ground states (Fig.~6) provides a direct explanation of this 
discrepancy, and namely, that the unit cells used in Ref.~\cite{Lemanski2} 
are too small to describe correctly the spin distributions.
 
Although we have presented here only the basic types of charge and 
magnetic superstructures 
they clearly demonstrate an ability of the model to describe different
types of charge and magnetic ordering. This opens an alternative route for 
understanding  of formation an inhomogeneous charge/magnetic order 
in strongly correlated electron systems.
In comparison to previous studies of this phenomenon based
on the Hubbard~\cite{Hubb} and $t-J$ model~\cite{t-J}, the study within the 
generalized spin-one-half FKM has one essential advantage 
and namely that it can be performed in a controllable way
(due to the condition $[f^+_{i\sigma}f_{i\sigma},H]=0$),
and in addition it allows easily to incorporate and examine
effects of various factors (e.g., an external magnetic field,
nonlocal interactions, etc.) on formation of charge and magnetic
superstructures. 

Of course, one can ask if these results persist also in the more realistic 
situations when additional interaction terms are included into the Hamiltonian 
(2). From the major interaction terms that come into account for the 
interacting $d$ and $f$ electron subsystems only the Hubbard type interaction
$U_{dd}\sum_i d^+_{i\uparrow}d_{i\uparrow}d^+_{i\downarrow}d_{i\downarrow}$ 
between the spin-up and spin-down $d$ electrons has been omitted in the 
Hamiltonian (2). In work~\cite{Lemanski2} Lemanski presents a simple justification for the 
omission of this term, based on an intuitive argument: the longer time electrons
occupy the same site, the more important becomes interaction between them. 
According to this rule the interaction between the itinerant $d$ electrons ($U_{dd}$)
 is smaller than the interaction between the localized $f$ electrons ($U_{ff}$) 
as well as smaller than the spin-independent interaction between the localized and 
itinerant electrons. Here we specify more precisely conditions when this 
term can be neglected. 
To determine the effects of $U_{dd}$ interaction on the ground-states 
of the spin-one-half FKM with spin-dependent interaction ($J=0.5$) in 
two-dimensions the 
exhaustive studies of the ground-state phase 
diagram of the model (in the $n_f-U_{dd}$ plane) have been  performed.
Of course, an inclusion of the $U_{dd}$ term makes the Hamiltonian (2) intractable 
by methods used for the conventional spin-one-half/spinless FKM and thus it was 
necessary
to use other numerical methods. Here we used the Lanczos method~\cite{Dagotto} to study exactly 
the ground states of the spin-one-half FKM generalized with $U_{dd}$ interaction 
between the spin-up and spin-down $d$ electrons. Such a procedure demands in 
practice a considerable amount of CPU time, which imposes severe restrictions 
on the size of clusters that can be studied within the exact-diagonalization method.
For this reason we were able to investigate exactly only the clusters up to $L=10$. 
The results of numerical calculations obtained for $U=4$ are summarized in Fig.~9
in the form of $n_f-U_{dd}$ phase diagram (the half-filled band case $n_f+n_d=1$ is 
considered). One can see that the ground-state configuration $w^0(N_f)$ found for 
$U_{dd}=0$ persists as a ground state up to relatively large values of $U_{dd}$ 
($U^c_{dd}\sim 3.5$), revealing small effects of the $U_{dd}$ term on the ground 
states of the model in the strong $U$ interaction limit. For this reason our 
numerical calculations have been done exclusively for large $U$. 

In conclusion, we have used a combination of small-cluster exact diagonalization 
calculations and a well-controlled numerical method to study the ground-state 
properties of the spin-one-half FKM extended by spin-dependent on-site 
interaction between localized and itinerant electrons in two dimensions. 
The results obtained have been used to construct the comprehensive picture of charge
and spin ordering in this model. It was shown that only a few 
configuration types form the basic structure of the charge phase diagram in the 
$N_f-N_d$ plane. In particular, the largest regions of stability correspond 
to the phase segregated configurations, the axial striped configurations and 
configurations that can be considered as mixtures of chessboard configurations 
and the full (empty) lattice. Moreover, it was 
found that the model exhibits a rich spectrum of magnetic solutions including 
various types of ferro- and antiferromagnetically ordered phases.

\vspace*{1cm}

This work was supported by Slovak Grant Agency VEGA under Grant
No.2/7057/27 and Slovak Research and Development Agency (APVV) under
Grant LPP-0047-06. H.C. acknowledges support of Stefan Schwartz
Foundation.

\newpage

%  Fig. 1
\begin{figure}[h]
\begin{center}
\includegraphics[width=12.5cm]{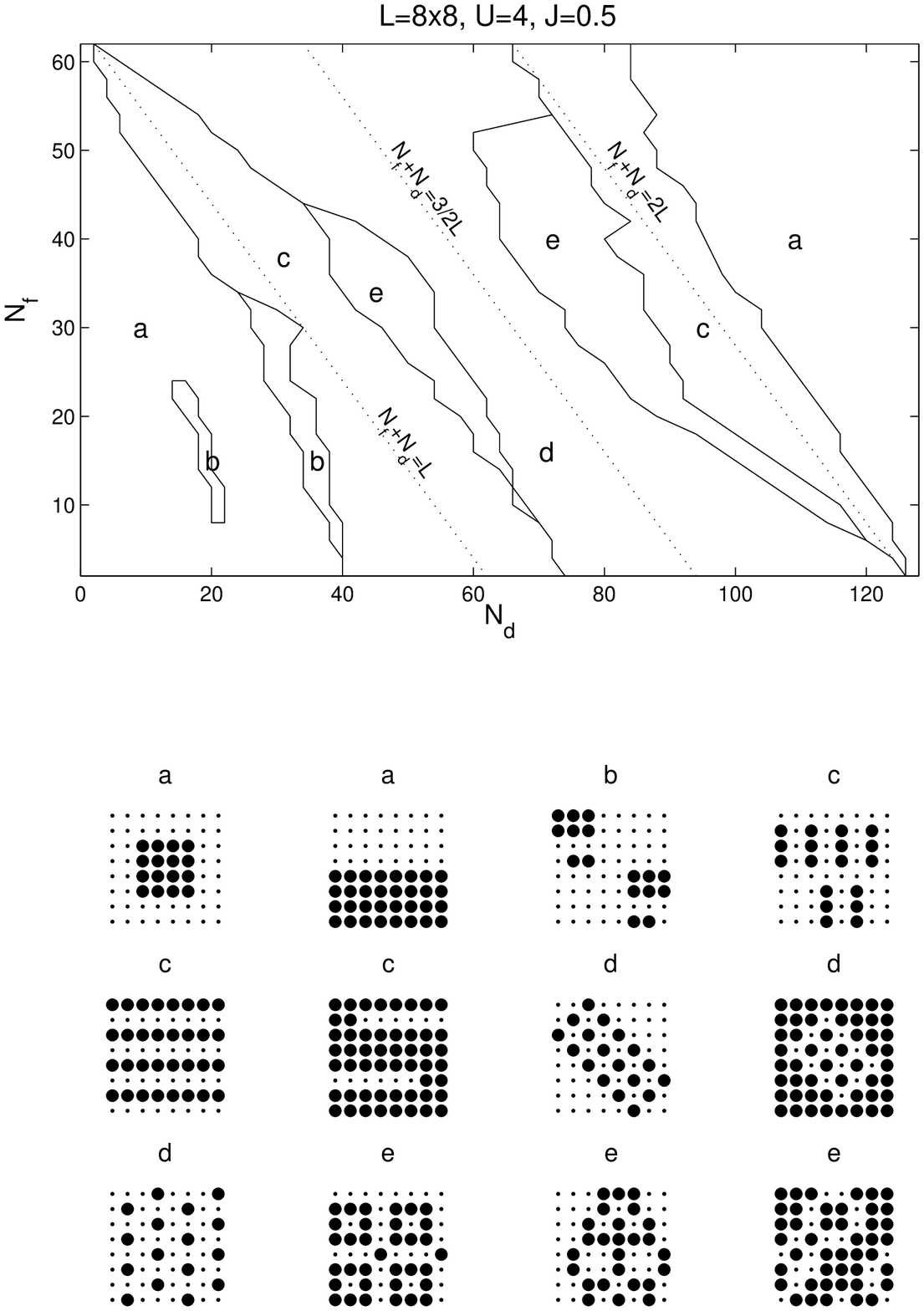}
\end{center}
\caption{The skeleton phase diagram of the two-dimensional spin-one-half FKM 
extended by  spin-dependent interaction calculated for $L=8\times 8$, $U=4$ 
and $J=0.5$.  (a) The segregated configurations, (b) the n-molecular phases, 
(c) the axial stripes, (d) the regular phases, the mixtures of regular phases 
and full/empty lattice and (e) the miscellaneous  
phases. Lower part: the typical examples of ground states of the model.
Large dots: sites occupied by $f$ electrons, small dots: vacant sites.}
\label{fig:1}
\end{figure}

%  Fig. 2
\begin{figure}[h]
\begin{center}
\includegraphics[width=13.5cm]{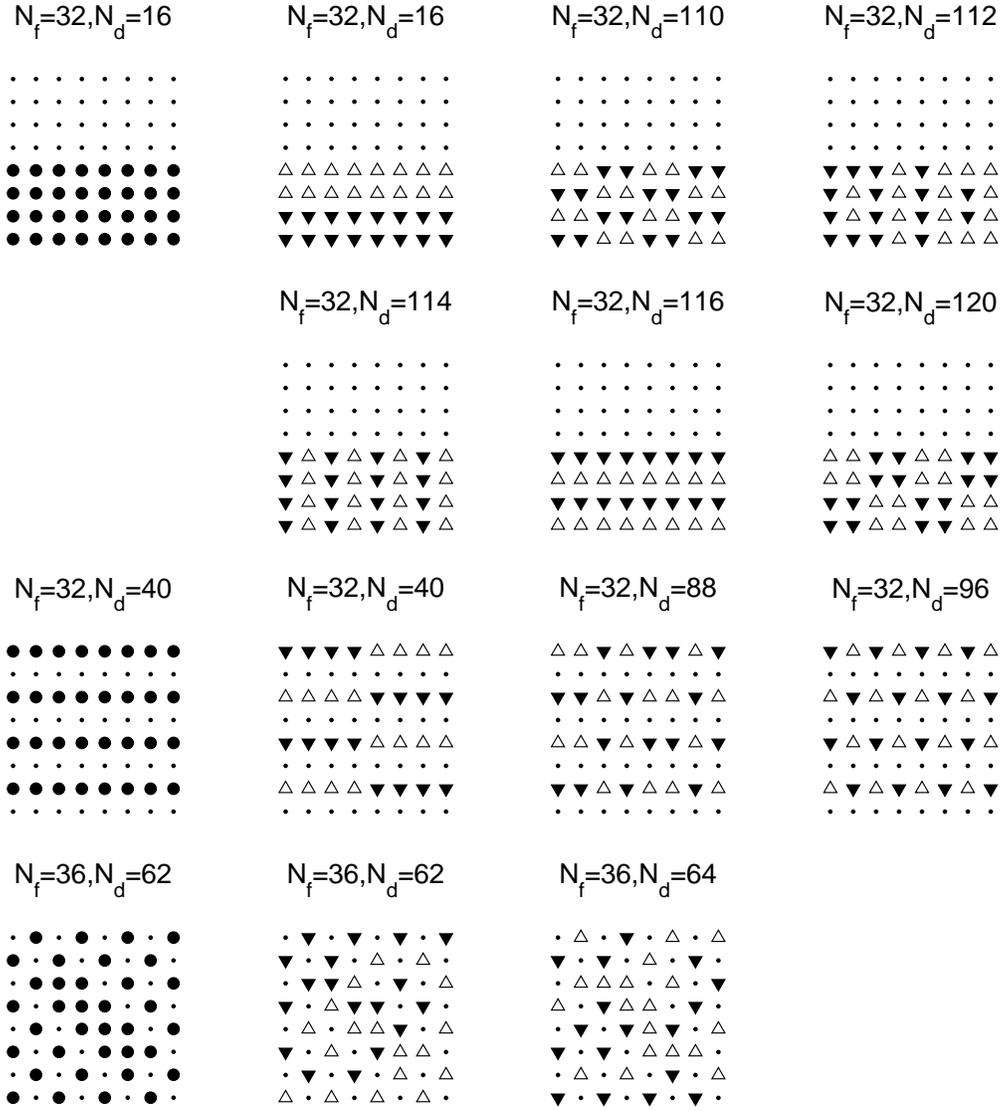}
\end{center}
\caption{Several spin distributions  with common charge pattern depicted on 
$L=8\times 8$ cluster. To visualize 
spin distributions we use $\bigtriangleup$ for the up spin electrons and 
$\bigtriangledown$ for the down spin electrons. }
\label{fig:1}
\end{figure}

%  Fig. 3
\begin{figure}[h]
\vspace*{-1cm}
\begin{center}
\includegraphics[width=15.0cm]{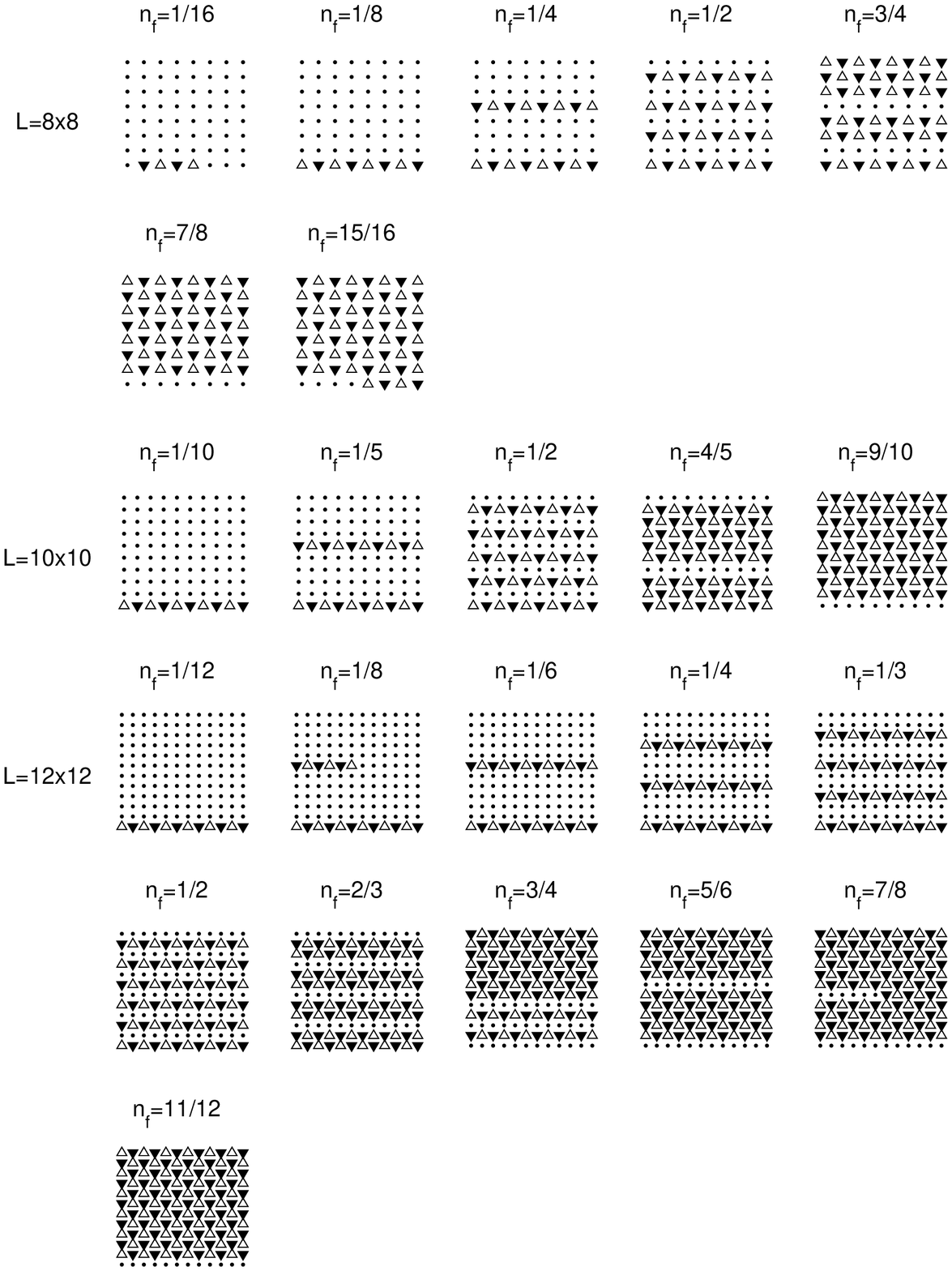}
\end{center}
\caption{Typical ground-state configurations of the two-dimensional generalized
FKM obtained for selected values of $n_f$ on finite clusters of $L=8\times 8$, 
$L=10\times 10$ and  $L=12\times 12$ sites at 
$U=4$, $J=0.5$ and $n_f+n_d=2$.}
\label{fig:1}
\end{figure}

%  Fig. 4
\begin{figure}[h]
\begin{center}
\includegraphics[width=11.0cm]{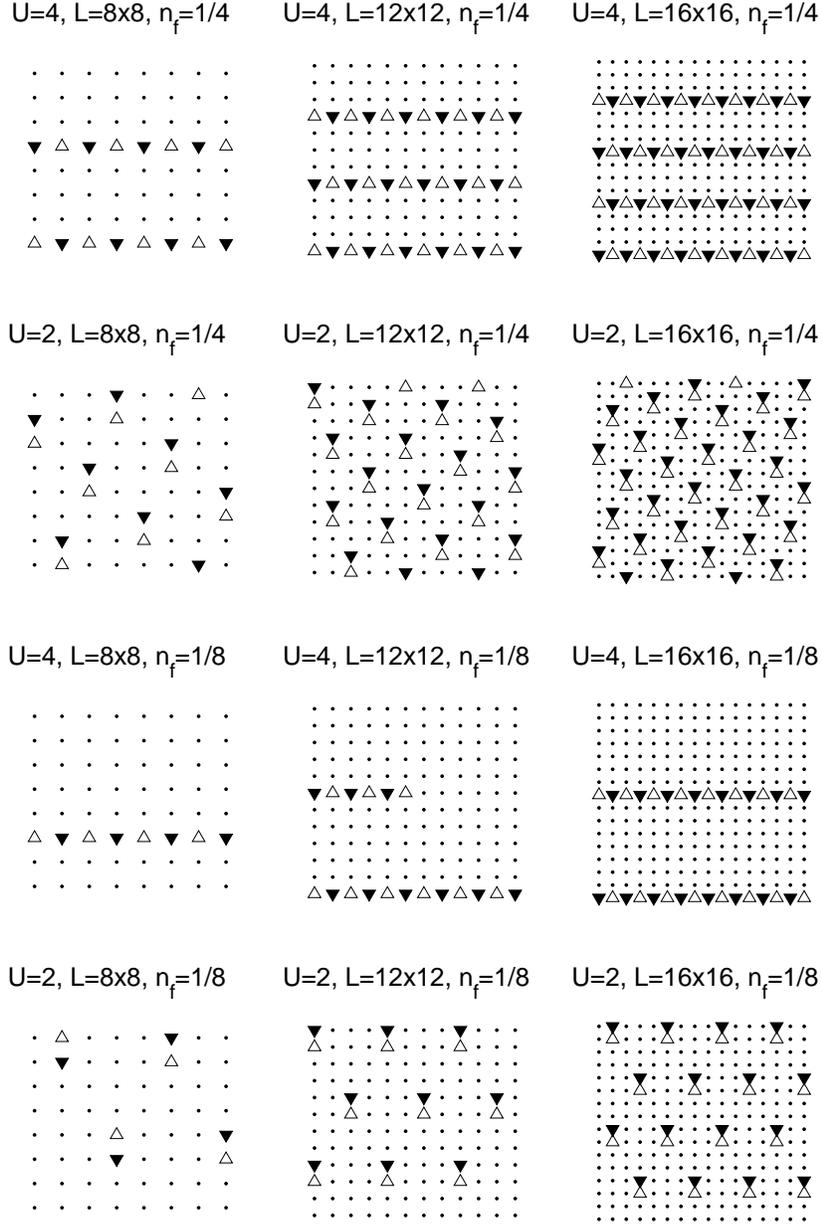}
\end{center}
\caption{Ground-state configurations of the two-dimensional generalized FKM 
obtained for different $U$ ($U=4$ and $U=2$) and $n_f$ ($n_f=1/4$, $n_f=1/8$ 
at $n_f+n_d=2$).}
\label{fig:1}
\end{figure}

%  Fig. 5
\begin{figure}[h]
\vspace*{-1cm}
\begin{center}
\includegraphics[width=15.0cm]{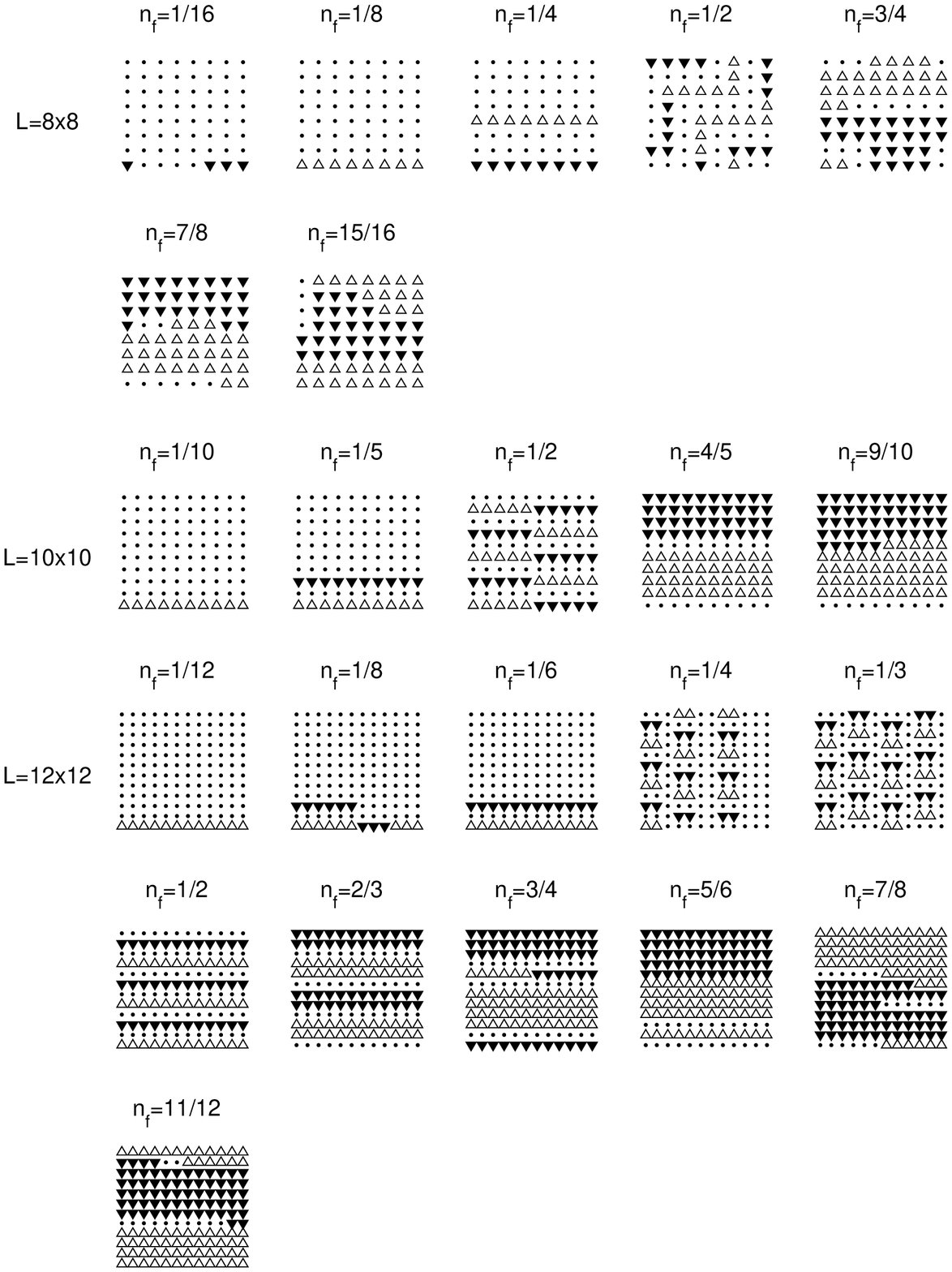}
\end{center}
\caption{
Typical ground-state configurations of the two-dimensional generalized
FKM obtained for selected values of $n_f$ on finite clusters of $L=8\times 8$, 
$L=10\times 10$ and  $L=12\times 12$ sites at 
$U=4$, $J=0.5$ and $n_f+n_d=1$.}
\label{fig:1}
\end{figure}

%  Fig. 6
\begin{figure}[h]
\vspace*{-1cm}
\begin{center}
\includegraphics[width=16.0cm]{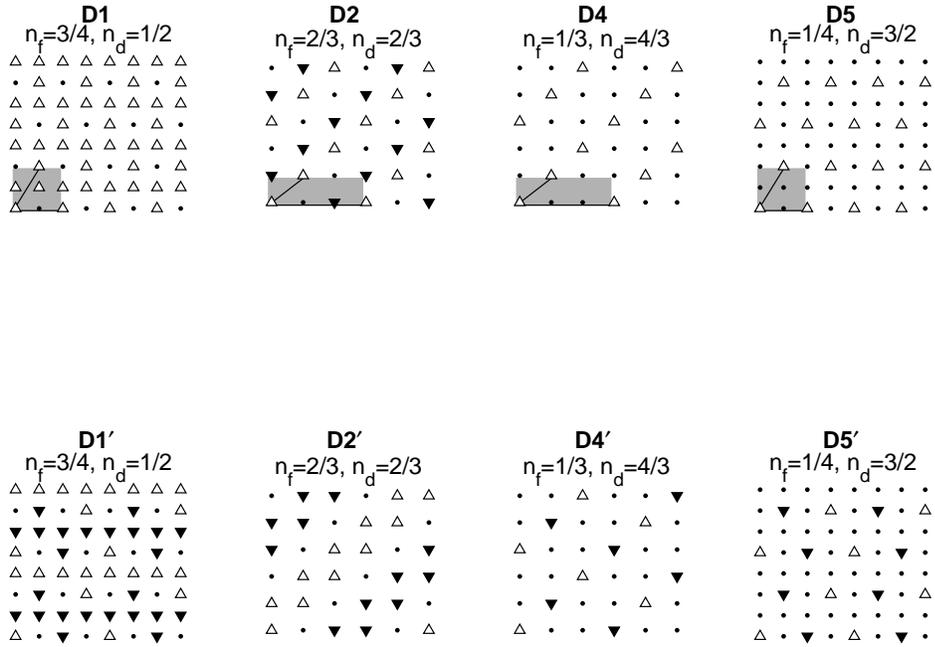}
\end{center}
\caption{The ground-state configurations of the two-dimensional spin-1/2 FKM 
for $U=8$ and four different pairs of $n_f$ and $n_d$ 
obtained by two different approaches: the method of restricted phase 
diagrams~\cite{Lemanski2} ($D1, D2, D4$ and $D5$) and our numerical method 
($D1', D2', D4'$ and $D5'$). The shaded region in the lower left corner shows 
the unit cell, and line segments show the translation vectors that are used to 
tile the two-dimensional plane.} 
\label{fig:1}
\end{figure}

%  Fig. 7
\begin{figure}[h]
\vspace*{-1cm}
%\hspace*{-4cm}
\begin{center}
\includegraphics[width=16.0cm]{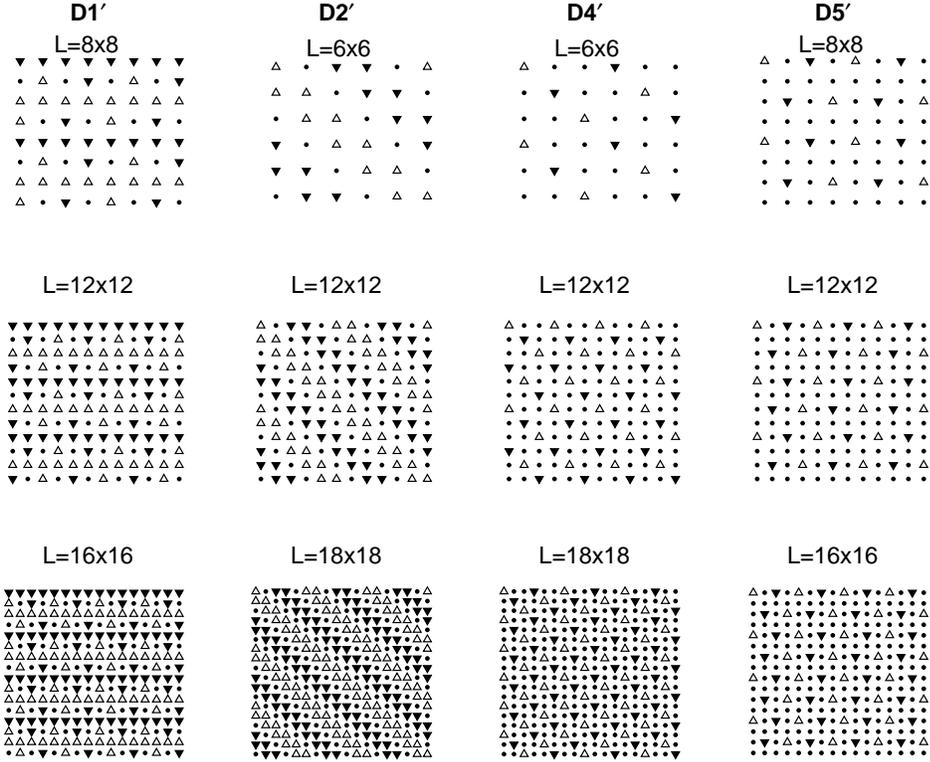}
\end{center}
\caption{The ground-state configurations of the two-dimensional spin-1/2 FKM 
calculated for the same $U, n_f$ and $n_d$ values as in Fig.~6 on 
different clusters of $L=8\times8$, $12\times12$ and $16\times16$ sites.}
\label{fig:1}
\end{figure}

%  Fig. 8
\begin{figure}[h]
\vspace*{-1cm}
\begin{center}
\includegraphics[width=15.0cm]{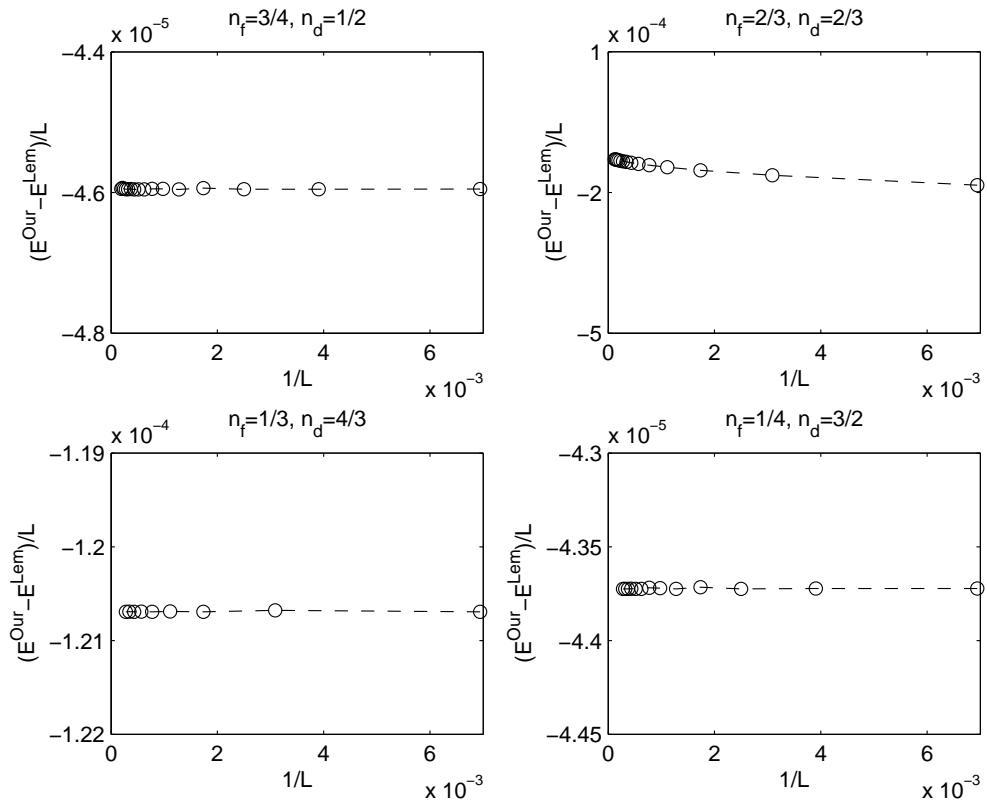}
\end{center}
\caption{The difference between our and the Lemanski's ground-state energies as 
a function of $1/L$ calculated for the extrapolated configurations corresponding 
to $D1', D2', D4', D5'$ and $D1, D2, D4, D5$ phases. The lines are only  
guides to the eye.}
\label{fig:1}
\end{figure}

%  Fig. 9
\begin{figure}[h]
\vspace*{-1cm}
\begin{center}
\includegraphics[width=15.0cm]{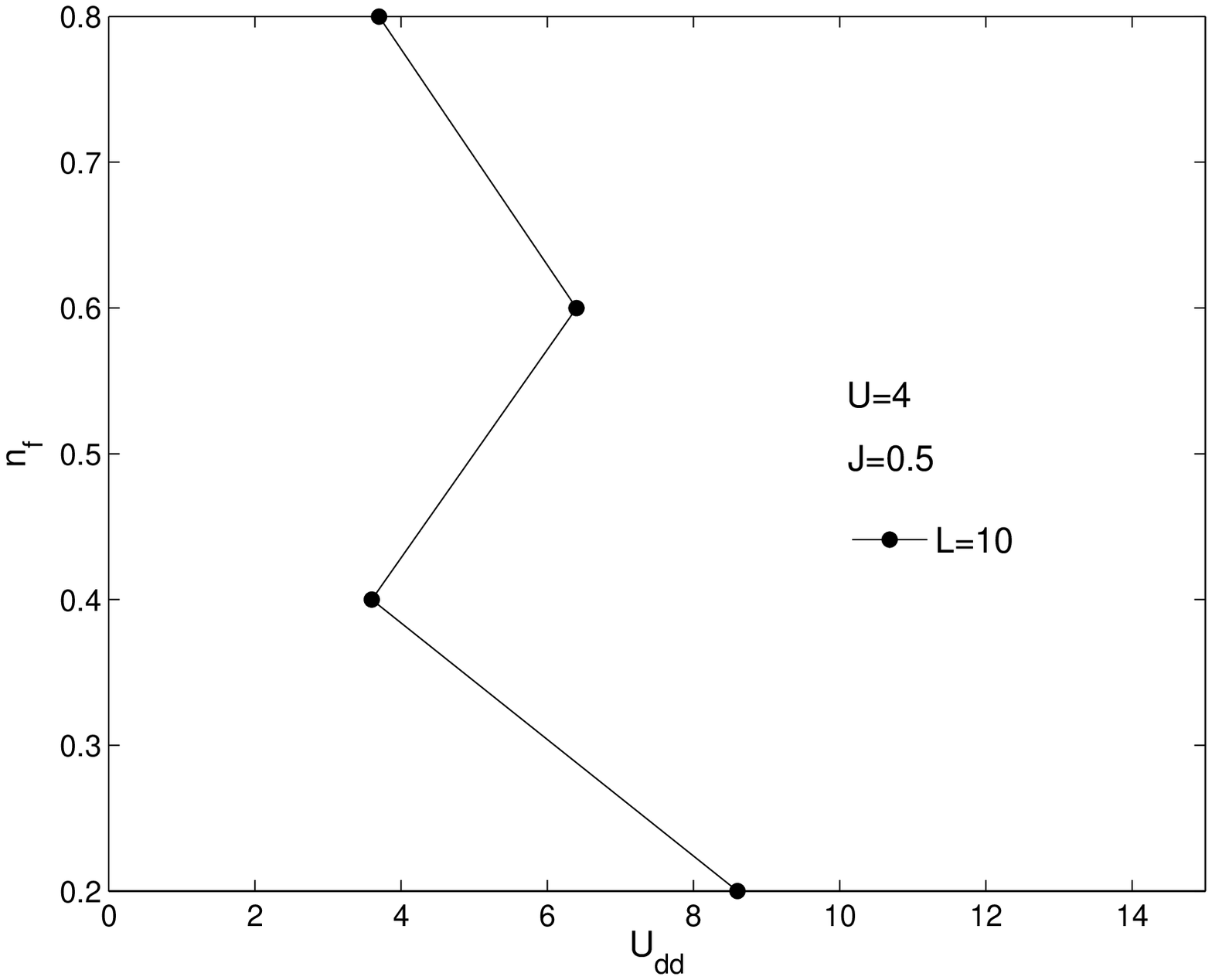}
\end{center}
\caption{The ground-state phase diagram of the spin-one-half FKM extended by the Hubbard 
interaction between the itinerant electrons calculated for $U=4$ and $J=0.5$ 
on small finite 
cluster of $L=10$ sites. Below $U^c_{dd}$ the ground states are the ground-state 
configurations of the conventional spin-one-half FKM ($U_{dd}=0$). Above $U^c_{dd}$ 
these ground states become unstable. The two-dimensional exact diagonalization 
results.}
\label{fig:1}
\end{figure}


\begin{thebibliography}{99}
\bibitem{Ni}C. H. Chen, S.-W. Cheong, and A. S. Cooper, Phys. Rev. Lett. {\bf 71},
2461 (1993); J. M. Tranquada, D. J. Buttrey, V. Sachan, J. E. Lorenzo, Phys. Rev. 
Lett. {\bf 73}, 1003 (1994); Phys. Rev. B {\bf 52}, 3581 (1995); V. Sachan, 
D. J. Buttrey, J. M. Tranquada, J. E. Lorenzo, G. Shirane, Phys. Rev. B 
{\bf 51}, 12742 (1995).

\bibitem{Cu} J. M. Tranquada, B. J. Sternlieb, J. D. Axe, Y. Nakamura, S. Uchida,
Nature (London) {\bf 375}, 561 (1995); Phys. Rev. B {\bf 54}, 7489 (1996); 
Phys. Rev. Lett. {\bf 78}, 338 (1997); 
H. A. Mook, P. Dai, and F. Dogan, Phys. Rev. Lett. {\bf 88}, 097004 (2002).

\bibitem{Co}I. Terasaki, Y. Sasago, and K. Uchinokura, Phys. Rev. B {\bf 56}, 
R12685 (1997); K. Takada, H. Sakurai, E. Takayama-Muromachi, F. Izumi, R. Dilanian, and 
T. Sasaki, Nature (London) {\bf 422}, 53 (2003).

\bibitem{FKM} L. M. Falicov, J. C. Kimball, Phys. Rev. Lett. {\bf 22}, 997 
(1969).

\bibitem{Lemanski} R. Lemanski, J. K. Freericks, G. Banach, Phys. Rev. Lett. 
{\bf 89}, 196403 (2002); J. Stat. Phys. {\bf 116}, 699 (2004).

\bibitem{Farky} P. Farka\v sovsk\'y, H. \v Cen\v carikov\'a, and N. 
Toma\v sovi\v cov\'a, Eur. Phys. J. B {\bf 45}, 479 (2005).

\bibitem{expspin} Y. Ando, K. Segawa, S. Komiya, and A. N. Lavrov, Phys. Rev. 
Lett. {\bf 88}, 137005 (2002); C. Howald, H. Eisaki, N. Kaneko, M. Greven, and 
A. Kapitulnik, Phys. Rev. B {\bf 67}, 014533 (2003).

\bibitem{Lemanski2} R. Lemanski, Phys. Rev. B {\bf 71}, 035107 (2005).

\bibitem{Farky1} P. Farka\v sovsk\'y, H. \v Cen\v carikov\'a, Eur. Phys. J. B 
{\bf 47}, 517 (2005).

\bibitem{Hubb} A. M. Oles, Acta Physica Polonica B {\bf 31}, 2963 (2000); J. Frohlich, 
D. Ueltschi, J. Stat. Phys. {\bf 118}, 973 (2005).

\bibitem{t-J} V. J. Emery, S. A. Kivelson, H. Q. Lin, Phys. Rev. Lett. 
{\bf 64}, 475 (1990); L. P. Pryadko, S. A. Kivelson, D. W. Hone, Phys. Rev. 
Lett. {\bf 80}, 5651 (1998); S. R. White, D. J. Scalapino, Phys. Rev. Lett. 
{\bf 80}, 1272 (1998); Phys. Rev. Lett. {\bf 81}, 3227 (1998); 
Phys. Rev. B {\bf 60}, R753 (1999); Phys. Rev. B {\bf 61}, 6320 (2000). 

\bibitem{Dagotto} E. Dagotto, Rev. Mod. Phys. {\bf 66}, 763 (1994).
\end{thebibliography}
\end{document}